# Plasmonic Route to Reconfigurable Polarization Optics

L. Li, T. Li,* X. M. Tang, S. M. Wang, Q. J. Wang, S. N. Zhu

National Laboratory of Solid State Microstructures, School of Physics, College of Engineering and Applied Sciences, Nanjing University, Nanjing, 210093

## Abstract

**Surface plasmon polariton (SPP) as a bounded mode on a metal/dielectric interface intrinsically has a definite transverse magnetic (TM) polarization that usually lacks further manipulations. However, the in-plane longitudinal components of SPP field can produce versatile polarization states when two orthogonal propagating SPP interfere with each other. Here, we demonstrated a plasmonic polarization router by designing appropriate nanohole arrays that can selectively scatter the interfered SPP fields to desired light beams. It is well proved that our device is able to reconfigure a certain input polarization to all kinds of states with respect to a scattered light. Accompanied with a composite phase modulation by diffractions, multiple focusing beams with different polarization states are simultaneously achieved, promising the possibility in polarization multiplexing and related signal processing. Our design offers a new route for achieving full control of the optical polarizations as well as the optical spin-orbital interactions.**

*Corresponding author: Tao LI,

Email: taoli@nju.edu.cn, URL: http://dsl.nju.edu.cn/litao/

Optical polarization is an important characteristic of light that carries rich information for signal processing in optical information technology. Compared with conventional optical elements, plasmonic devices provide more compact and efficient means in manipulating the polarization, for examples, plasmonic polarizers [1-5], polarization rotators and converters [6-10], and so on. Recently, plamsonic induced spin-orbital coupling has stirred new exciting ripples in the field of photonics [11-14], which is related with the polarization and phase tuning. In fact, the vectorial structure of plasmonic field gives rise to unique properties in the conversion of optical field between the radiation light and bounded surface plasmon polaritons (SPPs), where the polarization information of light can be reloaded by special SPP propagations in a controllable way [15-19]. However, most of these devices usually only offer limited functions in polarization control. To keep pace with the ever increasing requirement of information processing, a full controllable router to reconfigurable polarization states is what people are always in pursuit of.

In principle, the limitation of one or several polarization states in signal process can be overcome by a polarization router that has all polarization states at the same time if it does exist. Here, we demonstrate a plasmonic structure that can really stored all kinds of polarizations and route them selectively to preferred beams in a reconfigurable way. Thanks to the interference of longitudinal field of SPPs in our design, two orthogonally propagating SPPs surely construct such kind of near field bank with all polarization states, which have a regular distribution on metal surface. These bounded near field with distributed polarization information can be individually selected out by local scatterers on metal surface (e.g., nanohole). With an appropriate phase modulation on these scatterers, special light beams will be formed with a uniform polarization state. Based on this principle, plasmonic polarization routers were realized with selective focuses for different polarization states as required. As an impressive manifestation, eight focal spots by multiplexed focusing design are achieved simultaneously with eight kinds of polarization states (including linear, circular and elliptical states). As well as previous reports [17-19], this process can even be dynamically tuned by incident polarization.

The reconfiguration process is schematically shown in Fig. 1a. Two orthogonal propagating SPPs are launched from a polarized incidence by the L shaped slit on a metal film with the initial phase lag ($\Delta\varphi_0 = \varphi_x - \varphi_y$) between $x$ and $y$ directions. Different local phase lags of the two SPPs are reconfigured when they superposed at different positions of ($x, y$) as $\Delta\varphi = \Delta\varphi_0 + k_{SPP}(x - y)$, where $x$, $y$ are the distance to the initial slits in $x$ and $y$ directions respectively. The transverse components ($E_z$) of two orthogonal propagating SPPs, which are both normal to the metal surface, will interfere with each other resulting in a periodic intensity distribution along the diagonal direction. Whereas, the longitudinal components ($E_x$ and $E_y$) are orthogonal to each other, and will lead to a series of in-plane polarization states due to the field superposition with respect to different phase lags (see the inset of Fig. 1a). It is rightly due to this interference of longitudinal fields of SPP waves, all kinds of the polarization states can be built up simultaneously in a planar dimension, which do provide a polarization bank for further manipulations. By introducing nano-scatterers at proper locations, a special polarization will be selectively scattered out in a preferred manner if these scatterers phases are modulated at the same time. To be noted, the reconfigured in-plane field components ($E_x$ and $E_y$) would exactly converts to the scattered light beam while the normal component ($E_z$) does not [20], which ensures the efficiency of this polarization router.

In order to achieve constructive scattering to form a proper beam, phase relation of these scatterers should be carefully designed in despite of the routed polarization state, which is schematically shown in Fig. 1b and 1c. For examples, for an *x*-directional propagating SPP, the periodically arranged straight blue lines depicted in Fig. 1b correspond to a plane wave radiation, while Fig. 1c shows curved lines design indicating a focusing beam. Returning to the major point of the polarization routing, the red dashed diagonal lines corresponding to uniform polarization states have the interception of $\delta = n\lambda_{SPP} + \delta_0$ to *y*-axis, where *n* is an integer number, $\lambda_{SPP}$ is the wavelength of SPP, and $\delta_0$ is the initial intersection to *y*-axis. The nano-scatterers are designed on the intersections of the blue and red line, so that they will scatter out the fields with the same polarization state in a preferred manner. According to the correlated phase, the other branch of SPP in *y* direction will have the same phase modulation on these intersections, whose contribution has been included in the reconfigured polarizations.

In experiments, a certain polarized light, generated from a 633-nm laser light modulated by a polarizer and a quarter-wave plate, was focused by an objective lens (4×) and incident normally onto the sample from the substrate side. The scattered light was analyzed by a polarizer and a quarter-wave plate and then imaged by a CCD camera. Here, we set the fast axis of the incident and analyze quarter-wave plate as $\pi/4$ with respect to *x*-axis ($\alpha = -\beta = -\pi/4$, where $\alpha$ and $\beta$ are the angles of the fast axis to *x*-axis of the quarter-wave plates in the incident and transmitted sides respectively). The intensity from position (*x, y*) is proportion to $\cos^2(\theta_1 - \theta_2 + \Delta\varphi)$, where $\theta_1$ and $\theta_2$ are the angles of axis of incident and transmitted polarizer. A detailed deducing of the reconfigured polarization state of scattered beam with respect to the incidence polarization is provided in the Supplementary Section 1.

We first study the case of plane-wave scattering. It means the scattered beam have a plane phase front that can be achieved from a periodically arranged scatterers [21], as illustrated in Fig. 1b. The sample was fabricated by focused ion beam (FIB, Helios nanolab 600i, FEI) on 200 nm thick Ag film, which was sputtered on glass substrate. The L-shaped slit are 100 nm in width and penetrate through the Ag film for launching SPPs. The scatterers within the quadrant area are designed as nanoholes, which were milled with the diameter of 150 nm, depth of 80 nm and the period of 500 nm in *x*-direction as shown in Fig. 2a and 2b. The scattered light is recorded by CCD camera in the Fourier plane after a polarization analyzer. The Fourier image not only reflects a preferred k vector of the scattered light beam, but also carries a particular polarization state, which can be turned on or off on demand by the polarization analyzer. Figure 2c and 2d are the experimental results with respect to a defined nanohole array with its polarization state being selected and unselected by the analyzer, respectively. There is an apparent bright spot to be turned on (the upper one) and off (the lower one), see the zoom-in images in Fig. 2c and 2d. As has been interpreted as polarization router, the polarization state of the scattered beam is closely related to the location of the nanohole array, which correspond to different intersections of the phase line and polarization lines. All polarization states can be achieve by continuously shift the scatterer array.

In fact, focusing beams with routed polarizations should be of more practical usage. Our scattering approach does offer a convenient means to achieve the focuses, as schematically in Fig. 1c. The design details are provided in Supplementary Section 2. In order to achieve a symmetric result, one quadrant was extended to four, as shown in Fig. 3a. The structures in the left two quadrants (II and III) were designed with the initial intercept of $\delta_0 = \lambda_{SPP}/2$ targeting a focus at the position of (-7.5, 0, 40 μm) on the left side, while the right quadrants (I and IV) were designed for the right focus (7.5, 0, 40 μm) with $\delta_0 = 0$. To be noted, different intercepts corresponds to different polarization states in the scattering processes. So, if no polarization checked in analysis, two bright focal spots were found in the focal plane (z=40 μm) as shown in Fig. 3b. However, by adding a polarizer in analyses, two orthogonal states are clearly separated (see the spot in Fig. 3c for $\delta_0 = \lambda_{SPP}/2$, and Fig. 3d for $\delta_0 = 0$). It should be mentioned that the results of Fig. 3 are obtained with an right-circular polarization (RCP) incidence, and the detected focal spot in Fig. 3c revealed its orthogonal states (i.e., LCP), while the initial RCP state is maintained in Fig. 3d, according to the configuration of different intercept $\delta_0$. In fact, the intercept of the hole array rightly determine the polarization state routed for a required diffraction beam. That means, $\delta_0 = 0$ corresponds to the initial state, $\delta_0 = \lambda_{SPP}/2$ to the orthogonal one, and $\delta_0 = \lambda_{SPP}/4$ (and $3\lambda_{SPP}/4$) to the conversion from circular to linear polarizations. Therefore, taking good usage of the planar space, a two-fold polarization generator as well as the demultiplexer is achieved, which gives rise to what we emphasized - polarization router.

More importantly, thanks to the reconfigurable planar design with scatterers being capable of multiplexing, multiple desired polarization states in required beam-forms can be realized simultaneously. Here, to impressively manifest this function we demonstrate an eight-states-focuses router that is fulfilled by composite nanohole arrays with the spatial multiplexing. The structure was designed for focusing beams with focal spots to eight vertexes of a right octagon, whose initial intercepts have an equal gap of $\lambda_{SPP}/8$. The scattered light beams will be focused with eight different polarization states, which can be defined by the Poincaré sphere. In our experimental case, the incident polarization is RCP as well as the former one, saying, the amplitudes of $E_x$ and $E_y$ are equal, which determines that the designed eight polarization states will be on the great circle of Poincaré sphere (see Fig. 4c). The focuses are separated into 4 couples with the structures laid in the four quadrants in sequence. In each quadrant, two sets of arrays with initial intercept difference of $\lambda_{SPP}/2$ are designed and fabricated according to two orthogonal polarization states showing a mixed feature of the array, as shown in Fig. 4a. Under an RCP incidence, eight focuses are clearly recorded in the focal plane (z=40 μm) with similar intensities without polarization analyzer as shown in Fig. 4b, where the corresponding designed polarization states are sketched aside. To identify these states, the integrated intensities of the focuses with respect to different polarization analyzers (checked by the polarizer angles) are plotted in Fig. 4d, which shows good agreement with the theoretical predictions (see the corresponding curves in the figure). Here, two typical polarization states of routed focuses are demonstrated as shown in Fig. 4e and 4f, corresponding to the maximum in RCP-*a* and linear-*f* states (dashed lines (i) and (ii) in Fig. 4d), respectively. The intensities of the focuses are distinct especially for the orthogonal states (i.e., *a* and *b*, *e* and *f*), indicating the implementation of eight polarization states. As consequences, the simultaneous generation of multiple focuses with different polarization states on the great circle of the Poincaré sphere (Fig. 4c) is well proved. It should be mentioned other polarization states on the Poincaré sphere (e.g., on small circles) also can be achieved by tuning the amplitude ratio of $E_x/E_y$ of incidence, which is discussed in the Supplementary Section 1.

Now, our approach has demonstrated the powerful ability in routing the polarization and beam simultaneously. Actually, this strategy is quite general and adaptable. For instance, it can be adapted to an unidirectional SPP launching design proposed by Lin, et al [17], by which a spatial demultiplexer was additionally added in polarization routing, and a complete 2×2 polarization division multiplexer was developed (see the Supplementary Section 3). Moreover, this approach is independent to any resonance, which merits the whole process relatively low loss on one hand, and allows for a further alliance of the manipulation by the resonant design on the other hand. It means our method would provide a new dimension of freedom in manipulating SPP and light that can be coexistent with resonance modulation in the metasurface scheme [22,23] (e.g., every scatterer can be designed on and off- resonant for a further phase control). Furthermore, according to the results shown in Fig. 3 and Fig. 4, circular polarized lights are remitted well separately owing to the simultaneous polarization and phase modulation, which can be regarded as a kind of spin-orbital coupling for a optical spin Hall effect [11]. Then, our method probably offers a new scheme in designing new kinds of spin photonics with strong spin-orbital interaction [13].

In summary, we proposed and demonstrated plasmonic routers to reconfigure a certain incident polarization into various polarization states with desired beam forms. The critical routing process is established on the interference of orthogonal in-plane field of two perpendicularly propagating SPPs, which were launched by two crossed slits. Besides a polarization demultiplexing of two eigen states, eight polarization states were simultaneously generated and routed to eight focuses. Our strategy breaks the conventional means in polarization manipulations that were usually limited by the dual roles of plasmonic structures (i.e., both coupler and scatterer), and offers versatile and flexible opportunities in tailoring the polarization and phase of light all at once. Besides the revealed polarization division multiplexers, this approach is expected to promote the capability of people in full control of the light or photon and even open a new avenue in designing new kinds integrated functional photonic devices.

**Acknowledgement.** This work was supported by the National Key Projects for Basic Researches of China (No. 2012CB933501), the National Natural Science Foundation of China (Nos. 11174136, 11322439, 11321063, 91321312). This work was also supported by the Dengfeng Project B and Outstanding PhD candidate Program A of Nanjing University

**References**

1   Koerkamp, K. K., Enoch, S., Segerink, F., Van Hulst, N. & Kuipers, L. Strong influence of hole shape on extraordinary transmission through periodic arrays of subwavelength holes. *Phys Rev Lett* **92**, 183901 (2004).

2   Gordon, R. *et al.* Strong polarization in the optical transmission through elliptical nanohole arrays. *Phys Rev Lett* **92**, 037401 (2004).

3   Yu, N. *et al.* Semiconductor lasers with integrated plasmonic polarizers. *Appl Phys Lett* **94**, 151101 (2009).

4   Ellenbogen, T., Seo, K. & Crozier, K. B. Chromatic plasmonic polarizers for active visible color filtering and polarimetry. *Nano Lett* **12**, 1026-1031 (2012).

5   Wang, L. *et al.* Active display and encoding by integrated plasmonic polarizer on light-emitting-diode. *Scientific reports* **3**, 2603 (2013).

6   Zhang, J., Zhu, S., Chen, S., Lo, G.-Q. & Kwong, D.-L. An ultracompact surface plasmon polariton-effect-based polarization rotator. *Photonics Technology Letters, IEEE* **23**, 1606-1608 (2011).

7   Li, T., Wang, S., Cao, J., Liu, H. & Zhu, S. Cavity-involved plasmonic metamaterial for optical polarization conversion. *Appl Phys Lett* **97**, 261113 (2010).

8   Xu, J., Li, T., Lu, F., Wang, S. & Zhu, S. Manipulating optical polarization by stereo plasmonic structure. *Opt Express* **19**, 748-756 (2011).

9   Zhao, Y. & Alù, A. Manipulating light polarization with ultrathin plasmonic metasurfaces. *Phys Rev B* **84**, 205428 (2011).

10  Li, T. *et al.* Manipulating optical rotation in extraordinary transmission by hybrid plasmonic excitations. *Appl Phys Lett* **93**, 021110 (2008).

11  Shitrit, N., Bretner, I., Gorodetski, Y., Kleiner, V. & Hasman, E. Optical spin Hall effects in plasmonic chains. *Nano Lett* **11**, 2038-2042 (2011).

12  Shitrit, N. *et al.* Spin-optical metamaterial route to spin-controlled photonics. *Science* **340**, 724-726 (2013).

13  Yin, X., Ye, Z., Rho, J., Wang, Y. & Zhang, X. Photonic spin hall effect at metasurfaces. *Science* **339**, 1405-1407 (2013).

14  Li, G. *et al.* Spin-enabled plasmonic metasurfaces for manipulating orbital angular momentum of light. *Nano Lett* **13**, 4148-4151 (2013).

15  Genevet, P., Lin, J., Kats, M. A. & Capasso, F. Holographic detection of the orbital angular momentum of light with plasmonic photodiodes. *Nat Commun* **3**, 1278 (2012).

16  Gorodetski, Y., Drezet, A., Genet, C. & Ebbesen, T. W. Generating Far-Field Orbital Angular Momenta from Near-Field Optical Chirality. *Phys Rev Lett*

**110**, 203906 (2013).

17      Lin, J. *et al.* Polarization-controlled tunable directional coupling of surface plasmon polaritons. *Science* **340**, 331-334 (2013).

18      Rodríguez-Fortuño, F. J. *et al.* Near-field interference for the unidirectional excitation of electromagnetic guided modes. *Science* **340**, 328-330 (2013).

19      Miroshnichenko, A. E. & Kivshar, Y. S. Polarization Traffic Control for Surface Plasmons. *Science* **340**, 283-284 (2013).

20      Wang, J., Zhao, C. & Zhang, J. Does the leakage radiation profile mirror the intensity profile of surface plasmon polaritons? *Optics letters* **35**, 1944-1946 (2010).

21      Jun, Y. C., Huang, K. C. Y. & Brongersma, M. L. Plasmonic beaming and active control over fluorescent emission. *Nat Commun* **2**, 283(2011).

22      Yu, N. F. *et al.* Light Propagation with Phase Discontinuities: Generalized Laws of Reflection and Refraction. *Science* **334**, 333-337 (2011).

23      Ni, X., Emani, N. K., Kildishev, A. V., Boltasseva, A. & Shalaev, V. M. Broadband light bending with plasmonic nanoantennas. *Science* **335**, 427-427 (2012).

**Figure Captions**

Fig. 1. Reconfiguration of polarization states. (a) Schematic of the polarization reconfiguration process. The inset is the polarization states of in-plane SPP field

distribution with a RCP light incident. (b, c) Design strategy for routing the polarization states to special beams: plane wave (b) and focusing beam (c).

Fig. 2. Measurement of polarization properties. (a) and (b) Scanning electron microscope (SEM) image of the structure for plane wave radiation. *a* and *b* are the distances of the first nanocave from the vertical and horizontal slits. $\delta_0 = b - a$ is the designed initial intercept to the vertical slit. (c and d) Images on the Fourier plane with the polarization state selected (c) and unselected (d).

Fig. 3. Routing the reconfigured polarization states to focusing beams. (a) SEM image about the center of the nanostructure. (b) to (d) Focus images without polarization analyzing with a RCP incidence (b), analyzed with the polarization state of LCP (c) and RCP (d).

Fig. 4. Demonstration of simultaneous eight-states-focuses. (a) SEM image of the multiplexed nanostructure. (b) Focus image without polarization analyzing. The sketched signs are the corresponding polarization states of the focuses. (c) The polarization states of the focuses on the Poincaré sphere, in which the three Stokes parameters are $S_1 = |E_y|^2 - |E_x|^2$, $S_2 = 2|E_x||E_y|\sin\delta$, $S_3 = 2|E_x||E_y|\cos\delta$, defining the whole space of polarization states, and $\delta$ is the phase difference between the field components in the *x*- and *y*-axis. (d) Integrated intensities of the focuses at different polarization analyzing conditions. (e) and (f) Two focus images corresponding to the dash line (i) and (ii) in panel d.

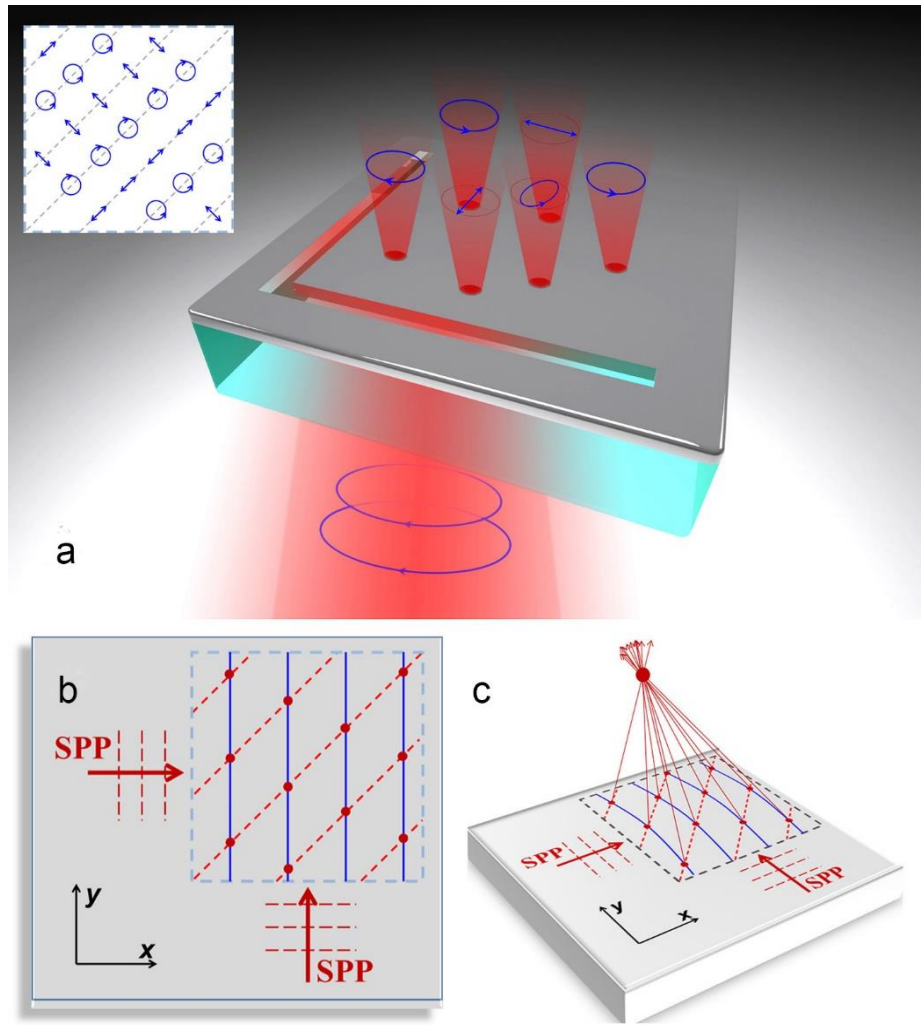

Fig. 1

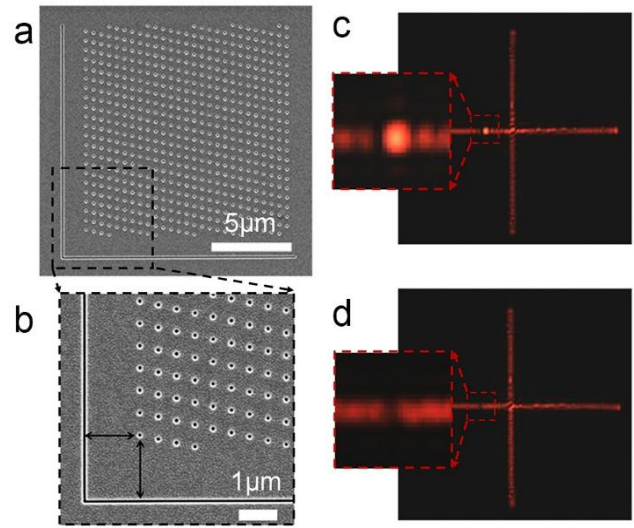

Fig. 2

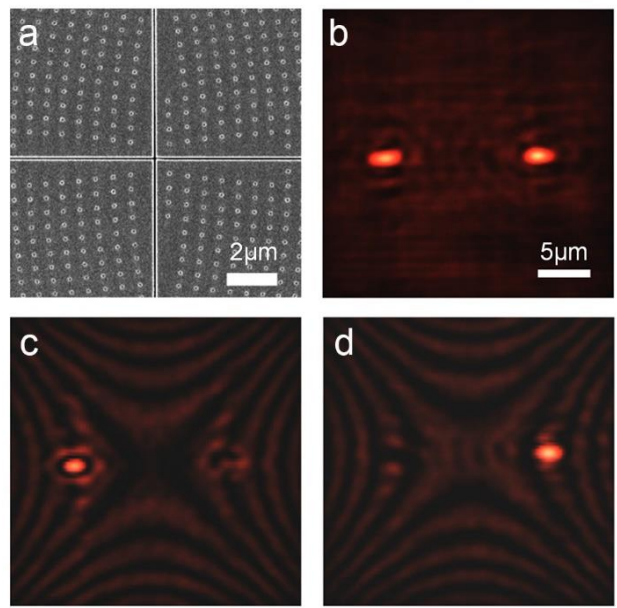

Fig. 3

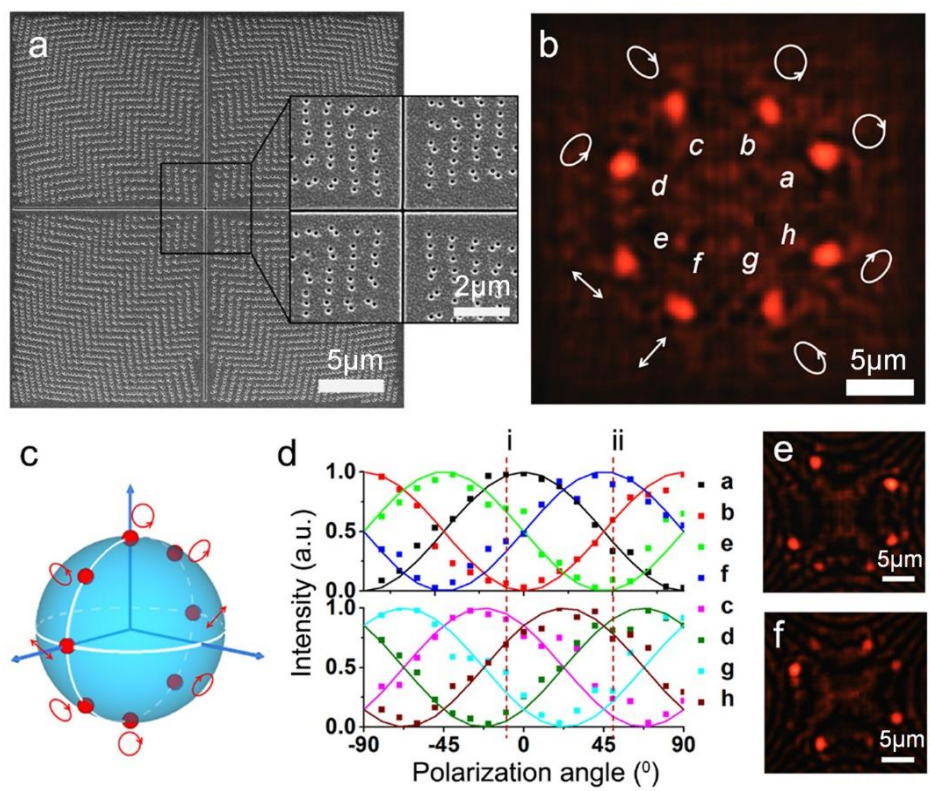

Fig. 4

# Supplementary information for

# Plasmonic Route to Reconfigurable Polarization Optics

L. Li, T. Li,* X. M. Tang, S. M. Wang, Q. J. Wang, S. N. Zhu

National Laboratory of Solid State Microstructures, School of Physics, College of Engineering and Applied Sciences, Nanjing University, Nanjing, 210093

## 1. Reconfiguring process of polarization states

A polarization state depends on the relative phases and the amplitudes of its two orthogonal components. In our polarization reconfiguring process, any polarization state on the Poincaré sphere can be synthesized by tuning the relative phases and amplitudes, which can be fulfilled by adjusting the initial intercepts of the array and the polarization angle of the incident light, respectively. Figure S1a is the optical system, where the horizontal and vertical slits in sample are set along the $x$- and $y$-directions respectively. The major axis of the incident polarizers and the fast axis of the quarter-wave plates have angles of $\theta_1$ and $\alpha$ to the $x$-axis, respectively, which are used to define a incident polarization. On the transmission side, another couple of polarizer and quarter-wave plate of ($\theta_2$ and $\beta$) are set as the polarization analyzer. The incident light modulated by the polarizer has a field of

$$A_0 = \begin{pmatrix} E_{H0} \\ E_{V0} \end{pmatrix} = \begin{pmatrix} \cos\theta_1 \\ \sin\theta_1 \end{pmatrix} E_0, \tag{1}$$

where $E_{H0}$ and $E_{V0}$ are the horizontal and vertical components of the modulated light. The polarized light is modulated by the incident quarter-wave plates to be $A_1 = G_1 \cdot A_0$, in which

$$G_1 = \begin{pmatrix} \cos^2\alpha + i\sin^2\alpha & (1-i)\sin\alpha\cos\alpha \\ (1-i)\sin\alpha\cos\alpha & \sin^2\alpha + i\cos^2\alpha \end{pmatrix}. \tag{2}$$

$A_1$ defines the relative amplitude and the initial phase lag $\Delta\varphi_0$ of the horizontal and vertical components. The two components will be coupled into SPPs propagating in $x$- and y-directions and have the phase of $k_{SPP}x$ and $k_{SPP}y$ at the position of (x, y) with respect to the vertical and horizontal slits, respectively. The field at this position is $A_2 = G_2 \cdot A_1$, in which,

$$G_2 = \begin{pmatrix} e^{ik_{SPP}x} & 0 \\ 0 & e^{ik_{SPP}y} \end{pmatrix} = e^{i(k_{SPP}x + k_{SPP}y)/2} \begin{pmatrix} e^{i\Delta\varphi_s/2} & 0 \\ 0 & e^{-i\Delta\varphi_s/2} \end{pmatrix} \quad (3)$$

and $k_{SPP}$ is the wave vector of SPPs at the Ag/air interface, $\Delta\varphi_s = k_{SPP}(x-y)$ is half of the phase difference of the two orthogonal propagating SPPs at the position of (x, y). It is this phase difference determined the reconfigured polarization state and the SPPs can be scattered out of the plane by nanostructures on the metal surface at the position with the polarization state.

The two components of the scattered light are modulated by the analyzing quarter-wave plate to be $A_3 = G_3 \cdot A_2$, where

$$G_3 = \begin{pmatrix} \cos^2\beta + i\sin^2\beta & (1-i)\sin\beta\cos\beta \\ (1-i)\sin\beta\cos\beta & \sin^2\beta + i\cos^2\beta \end{pmatrix}. \quad (4)$$

And finally, the light will be analyzed by another polarizer

$$G_4 = \begin{pmatrix} \cos\theta_2 & 0 \\ 0 & \sin\theta_2 \end{pmatrix}. \quad (5)$$

Therefore, the polarization state of scattered light at the local position of (x, y) is described as

$$A_4 = G_4 \cdot G_3 \cdot G_2 \cdot G_1 \cdot A_0. \quad (6)$$

Depending on this process, all kinds of polarization states can be synthesized and analyzed.

One typical example is in the case of that the amplitudes of the two components of the incidence are equal in x- and y-axis ($|E_x|^2 : |E_y|^2 = 1:1$), which can be performed by a circular polarized incidence ($\theta_1 - \alpha = \pm\pi/4$) or setting $\alpha = \pm\pi/4$. Based on the incidence, all polarization states on the great circle of the Poincaré sphere (Fig. S1b) can be synthesized by alter the initial intercepts. The corresponding analyzing condition and the final analyzed intensities are summarized in Table 1. To verifying it, we measured the integrated intensity of the Fourier point at different analyzing conditions with respect to 4 samples with different initial intercepts to the *y*-axis as the plan-wave scattering experiments in Fig. 2 of the main text. Fig. S1c depicts the results, where the symbols are experimental data and curves are the theoretical prediction, showing very good coincidence.

| Incidence | Analyzer | Intensity |
|---|---|---|
| $\alpha = \pi/4$ | $\beta = \pi/4$ | $\sin^2(\theta_1 + \theta_2 - \Delta\varphi_s/2)$ |
| $\alpha = \pi/4$ | $\beta = -\pi/4$ | $\cos^2(\theta_1 - \theta_2 - \Delta\varphi_s/2)$ |
| $\alpha = -\pi/4$ | $\beta = \pi/4$ | $\cos^2(\theta_1 - \theta_2 + \Delta\varphi_s/2)$ |
| $\alpha = -\pi/4$ | $\beta = -\pi/4$ | $\sin^2(\theta_1 + \theta_2 + \Delta\varphi_s/2)$ |
| $\theta_1 - \alpha = \pi/4$ | $\beta = \pi/4$ | $\cos^2(\theta_2 - \Delta\varphi_s/2)$ |
| $\theta_1 - \alpha = -\pi/4$ | $\beta = \pi/4$ | $\sin^2(\theta_2 - \Delta\varphi_s/2)$ |
| $\theta_1 - \alpha = \pi/4$ | $\beta = -\pi/4$ | $\sin^2(\theta_2 + \Delta\varphi_s/2)$ |
| $\theta_1 - \alpha = -\pi/4$ | $\beta = -\pi/4$ | $\cos^2(\theta_2 + \Delta\varphi_s/2)$ |

Table 1

It presents a clear statement of the polarization states reconfigures on the great circle of the Poincaré sphere. Indeed, any polarization state on other smaller circles of the Poincaré sphere can be synthesized as well. For instance, all of the polarization states on the smaller circle (Fig. S1b) corresponding to $|E_x|^2 : |E_y|^2 = 1:3$ can be obtained with linear polarized incident whose polarization is $60°$ to the *x*-axis. Likewise, the polar point (red spot on S1) on the Poincaré sphere in Fig. S1a can be realized by a linear polarized incidence with the polarization perpendicular to the *x*-axis as well. Other polarization states on different circle of the Poincaré sphere with relative amplitudes can be synthesized by different linear, elliptical or circular polarized incidence similarly. Various polarization states on the circle with required beam forms can be synthesized simultaneously with different initial intercepts under this incidence. The polarization states can be dynamically tuned by the relative amplitude and initial phase difference of the incident light.

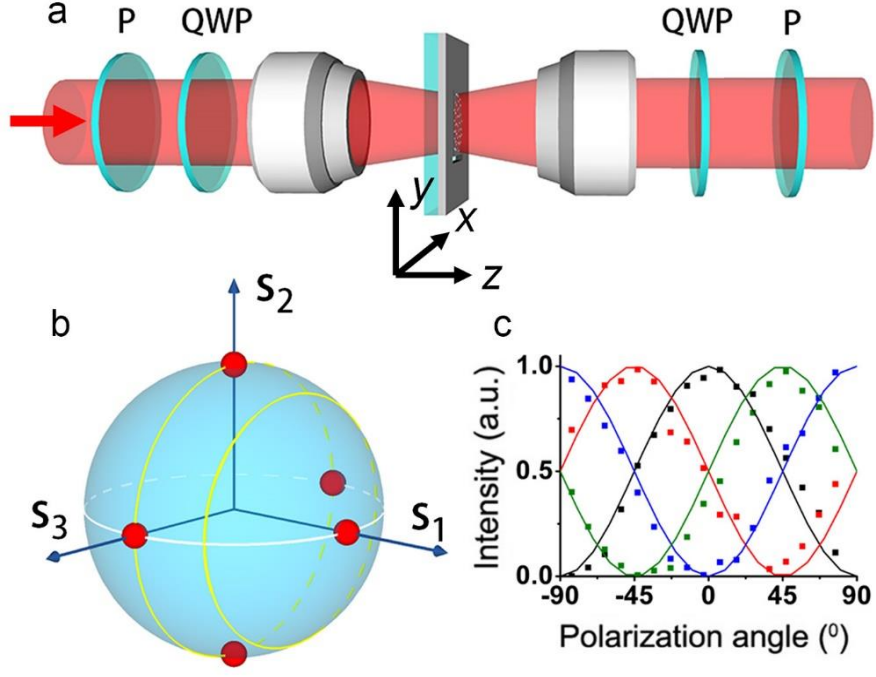

S1. (a) Optical system for experiment. (b) The reconfigured polarization states on the Poincaré sphere. The great circle is in the case of $\alpha = \pm \pi/4$ or circular polarized incidence so that $|E_x|^2:|E_y|^2 = 1:1$, and the smaller circle is in the case of $|E_x|^2:|E_y|^2 = 1:3$. (c) Integrated intensity of the spot for the plane-wave scattering on Fourier plane with different $\delta_0$. The blue, red, black and green lines are for $\delta_0 = \lambda_{SPP}/2, \lambda_{SPP}/4, 0, -\lambda_{SPP}/4$ respectively, corresponding for the four red spots on the great circle of the Poincaré sphere.

## 2. Design of focusing structures

The focus is designed with the interference principle, that is all the scattered lights from the structures $(x, y)$ have the same phase at the focal point of $(x_f, y_f, z_f)$ with respect to SPPs propagating in x-direction as:

$$k_{SPP}x + k_0 r = c \tag{7}$$

where $k_0$ and $k_{SPP}$ are the wave vector of light in free space and that of SPPs, $r = \sqrt{(x_f - x)^2 + (y_f - y)^2 + z_f^2})$ and $c$ is a constant. At the same time, the structure must satisfy

$$y = \pm x + n\lambda_{SPP} + \delta_0, \quad (8)$$

where $n$ is an integer and $\lambda_{SPP}$ is the wave length of SPP. The structures are obtained by solving the two equations together.

**3. 2×2 polarization division multiplexer by unidirectional SPP launcher**

It has been well demonstrated that eight different polarization has been achieved simultaneously. Whereas, like other polarization controller, there are only two orthogonal states can be totally separated or demultiplexed simultaneously in the scattering process. It is because the light only has two eigen states in polarization. However, SPP launching process does provide us a further freedom in multiplexing the polarization signals. According to a particular coupler design of aperture array proposed by Lin et al [17], SPP wave can be launched unidirectionally with circular polarized incidences. Therefore, it is quite appropriate to introduce into our four-quadrants device to replace the slit couplers, as schematically shown in Fig. S2a, where SPPs will be launched propagating to the quadrants II and IV with an RCP incidence, while to the other two quadrants for a LCP case. In order to achieve a demultiplexed focuses, quadrants II and IV are both designed with arrays for a left-up (-3, 3, 40 µm) and right-down focuses (3, -3, 40 µm) with intercepts of $\delta_0=0$ and $\delta_0=\lambda_{SPP}/2$ respectively, which manifests a composite arrays as shown in Fig. S2b. For an RCP incidence, the sub-array of $\delta_0=0$ in both quadrants II and IV will scatter the unidirectional SPPs into the left-up focus with a preserved RCP state, while the other ($\delta_0=\lambda_{SPP}/2$) to the right-down focus with a reconfigured LCP state. Similar function is also designed in quadrants II and IV with respect to an LCP incidence, which gives rise to the demultiplexed left-up and right-down focuses. Experimental results surely reproduced our theoretical predictions, as shown in Fig. S2d-S2g, with four well distinguished focusing spots. So that, a 2×2 polarization division multiplexer has been developed with a spatially selection SPP launching, in which two steps of polarization conversions from light to SPP and the reversal are efficiently taken usage of.

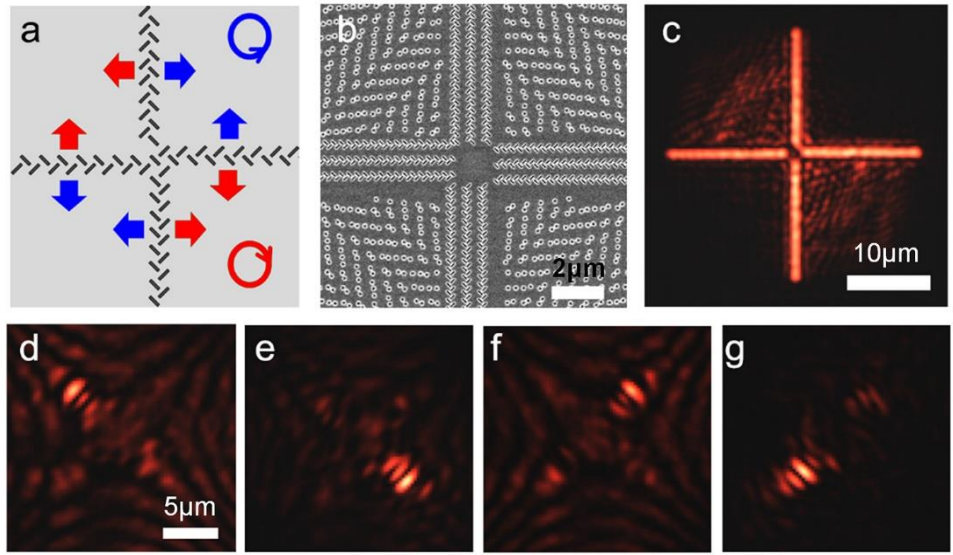

Fig. S2. Four focusing states for polarization demultiplexing. (a) Schematic for directional launching SPP. (b) SEM image of the center of the structure. (c) The image plane recorded by CCD camera with right circular polarization (RCP) incidence. (d) to (g) Different focus states for polarization routing: RCP incidence and analyzed in RCP with a selected $\delta_0=0$ (RCP→RCP by $\delta_0=0$) (d), RCP→LCP by $\delta_0=\lambda_{SPP}/2$ (e), LCP→LCP by $\delta_0=0$ (f), and LCP→RCP by $\delta_0=\lambda_{SPP}/2$ (g).

## 4. Scheme of possible application

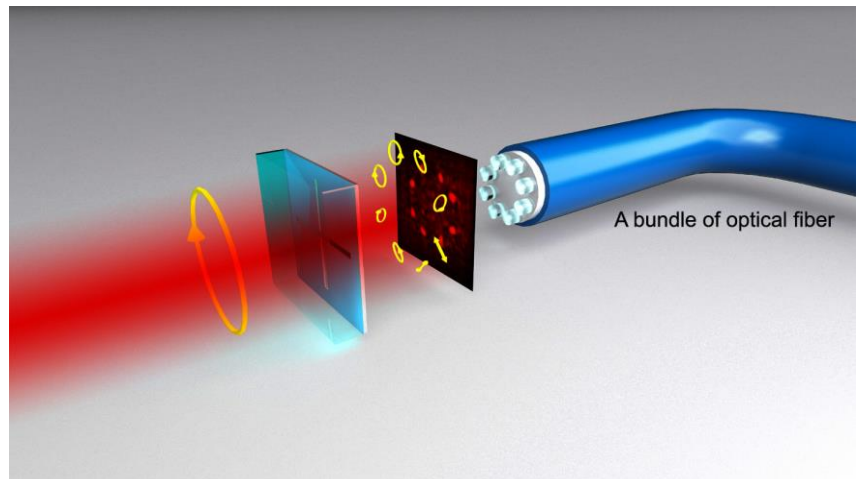

Fig. S3. Scheme of polarization router for a fiber communication with reconfigured eight polarization states.